\documentclass[twocolumn]{aastex631}
\usepackage{amsmath}
\usepackage{xspace}

\newcommand{\swift}{{\it Swift}\xspace}
\newcommand{\nicer}{\textit{NICER}\xspace}
\newcommand{\xmm}{{\it XMM-Newton}\xspace}
\newcommand{\chandra}{{\it Chandra}\xspace}
\newcommand{\cgs}{{\rm erg s$^{-1}$ cm$^{-2}$}\xspace}
\newcommand{\eros}{{eROSITA}\xspace}
\newcommand{\target}{{AT2018fyk}\xspace}

\expandafter\ifx\csname natexlab\endcsname\relax\fi
\providecommand{\url}[1]{\href{#1}{#1}}
\providecommand{\dodoi}[1]{doi:~\href{http://doi.org/#1}{\nolinkurl{#1}}}
\providecommand{\doeprint}[1]{\href{http://ascl.net/#1}{\nolinkurl{http://ascl.net/#1}}}
\providecommand{\doarXiv}[1]{\href{https://arxiv.org/abs/#1}{\nolinkurl{https://arxiv.org/abs/#1}}}

\shorttitle{AT2018fyk's second X-ray shutoff}
\shortauthors{Pasham et al.}

\graphicspath{{./}{figures/}}

\begin{document}

\title{A Potential Second Shutoff from AT2018fyk: An updated Orbital Ephemeris of the Surviving Star under the Repeating Partial Tidal Disruption Event Paradigm}

\author[0000-0003-1386-7861]{Dheeraj Pasham}
\affiliation{MIT Kavli Institute for Astrophysics and Space Research \\
		Cambridge, MA 02139, USA}

\author[0000-0003-3765-6401]{E.~R.~Coughlin}
\affiliation{Department of Physics, Syracuse University, Syracuse, NY 13210, USA}

\author[0000-0002-5063-0751]{M. Guolo}
\affiliation{Department of Physics and Astronomy, Johns Hopkins University, 3400 N. Charles St., Baltimore MD 21218, USA}

\author[0000-0002-0786-7307]{T. Wevers}
\affiliation{Space Telescope Science Institute, 3700 San Martin Drive, Baltimore, MD 21218, USA}

\author[0000-0002-2137-4146]{C.~J.~Nixon}
\affiliation{School of Physics and Astronomy, Sir William Henry Bragg Building, Woodhouse Ln., University of Leeds, Leeds LS2 9JT, UK}

\author[0000-0001-9668-2920]{Jason T. Hinkle}
\altaffiliation{NASA FINESST FI}
\affiliation{Institute for Astronomy, University of Hawai\`{}i at Manoa, 2680 Woodlawn Dr., Honolulu, HI 96822}

\author[0000-0003-3765-6401]{A.~Bandopadhyay}
\affiliation{Department of Physics, Syracuse University, Syracuse, NY 13210, USA}

\begin{abstract}
The tidal disruption event (TDE) \target showed a rapid dimming event 500 days after discovery, followed by a re-brightening roughly 700 days later. It has been hypothesized that this behavior results from a repeating partial TDE (rpTDE), such that prompt dimmings/shutoffs are coincident with the return of the star to pericenter and rebrightenings generated by the renewed supply of tidally stripped debris. This model predicted that 
the emission should shut off again around August of 2023. We report \target's continued X-ray and UV monitoring, which shows an X-ray (UV) drop in flux by a factor of 10 (5) over a span of two months, starting 14 Aug 2023. This sudden change can be interpreted as the second emission shutoff, which 1) strengthens the rpTDE scenario for \target, 2) allows us to constrain the orbital period to a more precise value of 1306$\pm$47 days, and 3) establishes that X-ray and UV/optical emission track the fallback rate onto this SMBH -- an often-made assumption that otherwise lacks observational verification -- and therefore the UV/optical lightcurve is powered predominantly by processes tied to X-rays. 
The second cutoff implies that another rebrightening should happen between May-Aug 2025, and if the star survived the second encounter, a third shutoff is predicted to occur between Jan-July 2027. Finally, low-level accretion from the less bound debris tail (which is completely unbound/does not contribute to accretion in a non-repeating TDE) can result in a faint X-ray plateau that could be detectable until the next rebrightening.  

\end{abstract}


\section{Introduction}
A tidal disruption event (TDE) occurs when a star approaches a supermassive black hole (SMBH) and is either completely or partially destroyed 
\citep[e.g.,][]{1988Natur.333..523R, gezari21}. TDE candidates were first discovered in the mid 1990s in the X-rays using the {\it ROSAT} soft X-ray telescope (e.g., \citealt{1996A&A...309L..35B, 1995A&A...300L..21G, 2002AJ....124.1308D}) and more recently with optical sky surveys like the ASASSN \citep{2014ApJ...788...48S}, ATLAS \citep{2018PASP..130f4505T}, Zwicky Transient Facility (ZTF; \citealt{ZTF}), and with {\it eROSITA} in the X-rays \citep{2021MNRAS.508.3820S}. With an estimated observed rate of roughly one TDE every 10$^{4-5}$ years per galaxy \citep{2023ApJ...955L...6Y, 2021MNRAS.508.3820S}, there is huge excitement for {\it Rubin} observatory (first light in 2024) which is expected to identify $>$100 events every year \citep{2011ApJ...741...73V,2020ApJ...890...73B}. 

A few dozens of TDEs are known so far and they have already transformed our understanding of SMBHs and their immediate surroundings. For example, some TDEs that were followed up extensively in the X-rays have shown powerful outflows 
(e.g., see \citealt{2024ApJ...963...75W, 2023ApJ...954..170K, 2024arXiv240112908A, 2018MNRAS.474.3593K}). Some systems have highly relativistic jets (bulk Lorentz factor $\sim$ a few tens) akin to blazars and have provided the best datasets to test models of jet launching (e.g., see \citealt{2023NatAs...7...88P, 2011Sci...333..203B, 2015MNRAS.452.4297B, 2015ApJ...805...68P, 2022Natur.612..430A, 2024ApJ...965...39Y}). In a few systems, radio synchrotron expanding at sub-relativistic speeds has been found which can be either from internal shocks within a jet \citep{2018ApJ...856....1P} or from external  shocks with ambient medium\citep{2023arXiv230813595C}. 

In addition to these TDE subclasses, in the last few years a surprising new sub-class has been uncovered: those that repeat on timescale of months to years \citep{2021ApJ...910..125P, 2022ApJ...926..142P, 2023ApJ...951..134P, wevers3, 2023A&A...669A..75L, evans23, guolo24, 2023arXiv231003782S}. These events have been postulated to arise from a star on a bound orbit about an SMBH that is partially disrupted during each pericenter passage. 
The TDE \target/ASASSN-18ul (redshift $z$=0.059, luminosity distance of 264.3 Mpc) is thought to be one example of this new class, and was discovered by the ASAS-SN optical survey in 2018, and follow-up monitoring with \swift, \nicer, \xmm, and \chandra showed that it remained X-ray and UV bright for roughly 500 days. Thereafter, it displayed a sudden and dramatic decrease in the X-ray (by a factor of $>$6000) and the UV (by a factor of $\approx$15; see Fig. \ref{Fig:fig2} and \citealt{wevers2}). The source also exhibited apparent state transitions similar to outbursting stellar-mass black hole binaries (soft/UV/accretion disk dominated state $\Rightarrow$ hard/X-ray/corona-dominated state $\Rightarrow$ quiescence; see \citealt{wevers2}). 

The source was then found to be X-ray and UV-bright \emph{again} around day $\sim 1200$$\footnote{All times in this paper are measured in observer's frame with respect to the optical discovery date of MJD 58369.2.}$, with \eros non-detections interspersed between the last non-detection at day 600 and the first new detection at day 1200, showing that \target suddenly ``turned on'' following a $\sim 2$ year dark period -- behavior that is otherwise unprecedented in observed TDEs. The precipitous drop in luminosity and the rebrightening can be explained by the rpTDE scenario$\footnote{The presence of AGN in \target was ruled out based on detailed analyses  of multi-wavelength data of the host galaxy, see section 2.4 of \citet{wevers3}}$: \cite{wevers3} argued that if the return of the tidally disrupted debris to the SMBH is tightly coupled to the accretion rate and the corresponding luminosity, which is a good approximation for highly relativistic settings with small viscous delays, the sudden cessation of accretion coincides with the return of the star to pericenter, and the time between the sudden cutoff and the rebrightening equates to the fallback time of the tidally stripped debris. With this model, they deduced that the orbital period of the star is $\sim 1200$ days. They predicted that, if the star was not destroyed during the second pericenter passage, the system should display another dimming in August 2023, analogous to the one observed in 2019.

Here, using continued X-ray monitoring with \swift, \nicer, \xmm and \chandra, we report the finding of this second cutoff at 1830$\pm$29 days (14 Aug - 11 Oct 2023). Our data analysis is shown in section \ref{sec:data} while we discuss the implications and provide specific predictions to further test the rpTDE model in section \ref{sec:dis}.


\section{Data and Analysis}\label{sec:data}
We used the following cosmological parameters: $\Lambda$CDM cosmology with parameters H$_{0}$ = 67.4 km s$^{-1}$ Mpc$^{-1}$, $\Omega_{\rm m}$ = 0.315 and $\Omega_{\rm \Lambda}$ = 1 - $\Omega_{\rm m}$ = 0.685 \citep{2020A&A...641A...6P}.

\subsection{Swift X-Ray Data}\label{sec:swift}
\swift \citep{swift} observed \target on 210 occasions as of 5 Jan 2024. Out of these, 5 were corrupted or did not have Photon Counting (PC) data and were excluded. Observations up to MJD 59809, i.e., 178 of these observations, were reported in \cite{wevers3}. Here we present additional monitoring data since 18 August 2022. For consistency, we reduce the entire \swift archival data of \target here. 

We started our analysis by downloading the data from HEASARC public archive (\url{https://heasarc.gsfc.nasa.gov/cgi-bin/W3Browse/w3browse.pl}) and reduced the X-Ray Telescope (XRT; \citealt{xrt}) observations on a per ObsID basis using the HEASoft tool {\tt xrtpipeline}. Then we extracted source and background count rates in the 0.3-10.0 keV band using the {\it ftool} {\tt xrtlccorr}. We used a circular aperture of radius 47$^{\prime\prime}$ for source and an annulus of inner and outer radii of 70$^{\prime\prime}$ and 235$^{\prime\prime}$, respectively. These values were chosen to ensure there are no contaminating sources within the chosen boundaries. From these, we obtained a net (background-subtracted) rate for each ObsID.

\target was especially faint with net rates close to zero in the most recent observing campaign since MJD 60000 (approved \swift cycle 19  program 1922148; PI: Pasham). Therefore, we carefully analyzed them by stacking them into 4 datasets with the following time boundaries: MJD 60030-60070 (L1), MJD 60070-60140 (L2), MJD 60140-60200 (L3), and MJD 60200-60310 (L4). The source was detected in two of these four stacked datasets. The net 0.3-10.0 keV count rate/3$\sigma$ upper limit for L1, L2, L3 and L4 epochs were (1.5$\pm$0.6)$\times$10$^{-3}$ cps, $<$4.3$\times$10$^{-3}$ cps, (2.6$\pm$0.7)$\times$10$^{-3}$ cps, and $<$2$\times$10$^{-3}$ cps, respectively. We also visually inspected the exposure-corrected 0.3-10.0 keV image for epoch L3 in which a point source is evident (see the middle panel of Fig. \ref{Fig:fig1}). Assuming a spectrum similar to the one implied by an \xmm observation taken around that time, the flux conversion factor is 3.1$\times$10$^{-11}$  \cgs/counts sec$^{-1}$.

\subsection{Swift UV Data}\label{sec:uvot}
UV observations were taken with \swift/UVOT contemporaneously with the XRT observations. 
We used the \texttt{uvotsource} package to measure the UV photometry, using an aperture of 5\arcsec. We subtracted the host galaxy contribution by modeling archival photometry data with stellar population synthesis using \textsc{Prospector} \citep{Johnson_21}, following the procedure described in \citet{wevers2} and tabulated in their table 2. We apply Galactic extinction correction to all bands using $E(B-V)$ value of 0.011 from \citet{Schlafly2011}. 

\subsection{XMM-Newton}\label{sec:xmm}
\xmm observed \target on 8 occasions (ObsIDs: 0831790201, 0853980201, 0854591401, 0911790701, 0911790601, 0911791501, 0911791401 and 0921510101). Two observations (0911790701 and 0911791501) did not have any science data and the rest, except for the latest one (ObsID: 0921510101), have been published elsewhere \citep{wevers1,wevers2,wevers3}. This latest dataset was part of an approved \xmm cycle 22 Guest Observer Target Of Opportunity (GO ToO program 92151; PI: Pasham) to capture the second X-ray shutoff of \target. While the main focus in this work will be on this latest dataset we also reduce all the others here for uniformity. 

We started \xmm data analysis by downloading the data from the HEASARC public archive (\url{https://heasarc.gsfc.nasa.gov/cgi-bin/W3Browse/w3browse.pl}). Then we ran the {\tt epproc} tool of XMMSAS software to reduce the European Photon Imaging Camera (EPIC)’s pn detector. We did not use MOS data in this work. First, we visually inspected the background in all the six ObsIDs following the steps outlined in the data analysis thread: \url{https://www.cosmos.esa.int/web/xmm-newton/sas-thread-epic-filterbackground-in-python}. All observations were affected by background flares to some extent and we removed those epochs to obtain a set of Good Time Intervals (GTIs) per ObsID. Source events were extracted from a circular aperture with a radius of 30$^{\prime\prime}$ while background events were extracted from a nearby circular aperture free of any point sources with a radius of 50$^{\prime\prime}$. The source is clearly detected in all but 0854591401 (XMM3 as per \citealt{wevers3}). Consequently, five spectra were extracted following the standard procedure outlined here: \url{https://www.cosmos.esa.int/web/xmm-newton/sas-thread-pn-spectrum}. The spectra were grouped using the {\tt specgroup} task of \xmm software (XMMSAS) to have minimum of 1 count per spectral bin. Cash statistic was used for spectral modeling in {\it XSPEC} \citep{xspec}. For each spectrum, we only used the bandpass where the source is above the background (see Table \ref{table}).

The most recent dataset is consistent with a simple powerlaw modified by MilkyWay absorption of 1.2$\times$10$^{20}$ cm$^{-2}$ (C-stat/degrees of freedom (dof) of 110/114). Additional absorption at the host redshift is not required by the data in any of the five spectra. The best-fit powerlaw index in the most recent dataset is  1.96$^{+0.86}_{-0.88}$ (see Table \ref{table} for details on flux and luminosity). The spectrum did not have enough signal-to-noise to test more complicated spectral models. 


\begin{figure*}[ht]
    \includegraphics[width=\textwidth]{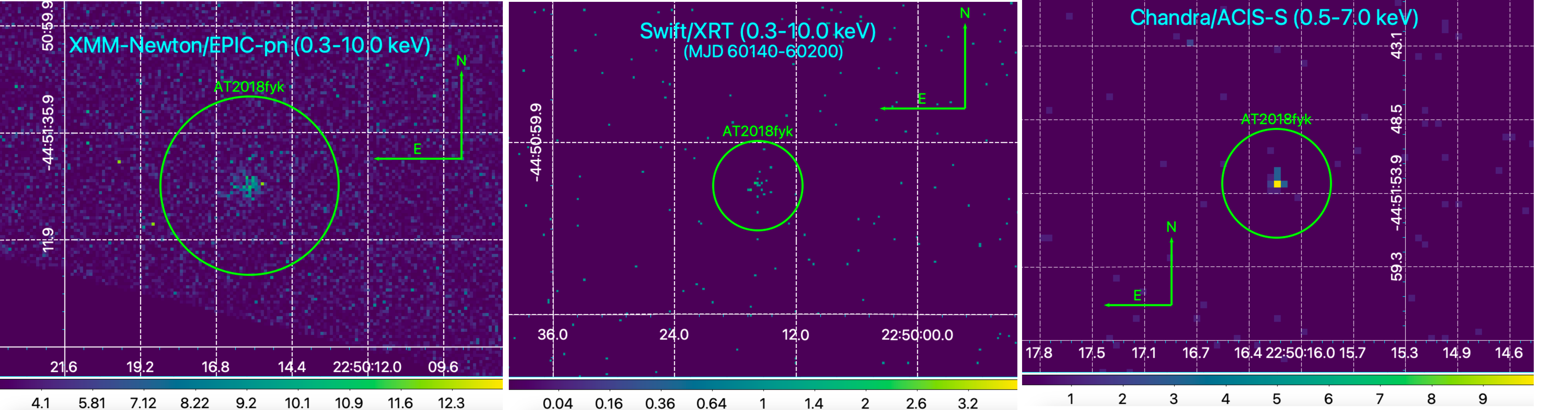}
    \caption{ {\bf Left:} \xmm/EPIC-pn image of \target's field of view on MJD 60102.76 (XMM ObsID 0921510101). The circle centered on \target has a radius of 30$^{\prime\prime}$ while the arrows pointing North and East are each 30$^{\prime\prime}$ in length. {\bf Middle:} Stacked \swift/XRT image of \target using data taken between MJDs 60140 and 60200, i.e., the data point around day 1800 between \xmm (pentagon) and \chandra detections (square) in Fig. \ref{Fig:fig2} The statistical significance of the detection is 3.7$\sigma$. The circle has a radius of 47$^{\prime\prime}$ while the directional arrows are 90$^{\prime\prime}$ each. {\bf Right:} Stacked \chandra/ACIS X-ray image using data from ObsIDs 28294 and 28972. The circle centered on \target has a  4$^{\prime\prime}$ radius while the directional arrows are 5$^{\prime\prime}$ each.}
    \label{Fig:fig1}
\end{figure*}

\subsection{Chandra}\label{sec:chandra}
\chandra’s Advanced CCD Imaging Spectrometer (ACIS) observed \target on three occasions: MJD 59029.22 (29 June 2020), MJD 60227.71 (10 October 2023; ObsID: 28294) and MJD 60228.58 (11 October 2023; ObsID: 28972). All these were carried out in the ACIS-S array mode and we use the nominal bandpass of 0.5-7.0 keV throughout. The first observation was published in \cite{wevers2} while the most recent two datasets were observed as part of an approved \chandra Cycle 25 guest observer program to catch the source during the second shutoff phase predicted by \cite{wevers3} (PI: Pasham; GO proposal number 25700383). For consistency, we reduce all the three datasets here. 

We started our data analysis by reducing the data with the {\tt chandra\_repro} tool of CIAO 4.16 software using the latest CALDB 4.11.0. First, we extracted exposure-corrected images in the 0.5-7.0 keV bandpass using the {\tt fluximage} task of CIAO and see an excess at the position of \target in both of the most recent observations (IDs: 28294 and 28972). Next, we extracted the X-ray spectra and relevant response files for each of the two recent observations separately using {\tt specextract} tool of CIAO. These spectra were grouped to have a minimum of 1 spectral count per bin using the {\it optmin} flag of the HEASoft {\it ftool} {\tt ftgrouppha}. We modeled them together in {\it XSPEC} \citep{xspec} with a powerlaw model modified by MilkyWay neutral absorption column of 1.2$\times$10$^{20}$ cm$^{-2}$ ({\it tbabs*zashift*pow}). With only 25 net (background-corrected) counts the spectral index is poorly constrained. Therefore, we fixed it at the best-fit value from the \xmm data of 1.96. This yields a best-fit C-statistic/dof of 38.5/56 and an observed 0.3-10.0 keV flux (luminosity) of (9.0$^{+1.0}_{-2.0}$)$\times$10$^{-15}$ \cgs (7.0$^{+2.0}_{-1.0}$$\times$10$^{40}$ erg s$^{-1}$). This represents a factor of $>$7 decrease in flux since the latest \xmm observation taken roughly 4 months earlier.   

\subsubsection{\chandra astrometry}
We also combined the two observations to estimate an accurate position by following the steps outlined in \url{https://cxc.cfa.harvard.edu/ciao/threads/fluxes_multiobi/}. We computed the offsets between the two datasets to be 0.18 pixels and 0.42 pixels in the X and Y directions, respectively. To improve this we performed astrometric correction to obsID 28972 to match with that of 28294 which has about 60\% higher exposure time (33 ks vs 20 ks). Following the steps outlined in the above \chandra data analysis thread we reduced the offsets to 0.15 pixels and 0.06 pixels, respectively. An X-ray (0.5-7.0 keV) image from combining obsIDs 28294 and 28972 is shown in the right  panel of Fig. \ref{Fig:fig1}. The source region defined as a circular aperture of 4$^{\prime\prime}$ in radius has 25 net counts. Running {\tt wavdetect} on this combined images yields a best-fit X-ray position of (22:50:16.17,-44:51:53.00) with an uncertainty of 0.12$^{\prime\prime}$ in each direction. This is consistent with the best-fit {\it Gaia} position reported in \cite{wevers1} based on the optical emission during the first outburst in 2018. 

\subsection{Hubble Space Telescope (HST)}
The UV measurement from Hubble Space Telescope's {\it F275W} filter with an effective wavelength of 2750 \AA~ was taken from \cite{wen24}.

\begin{figure*}
    \centering
    \includegraphics[width=\textwidth]{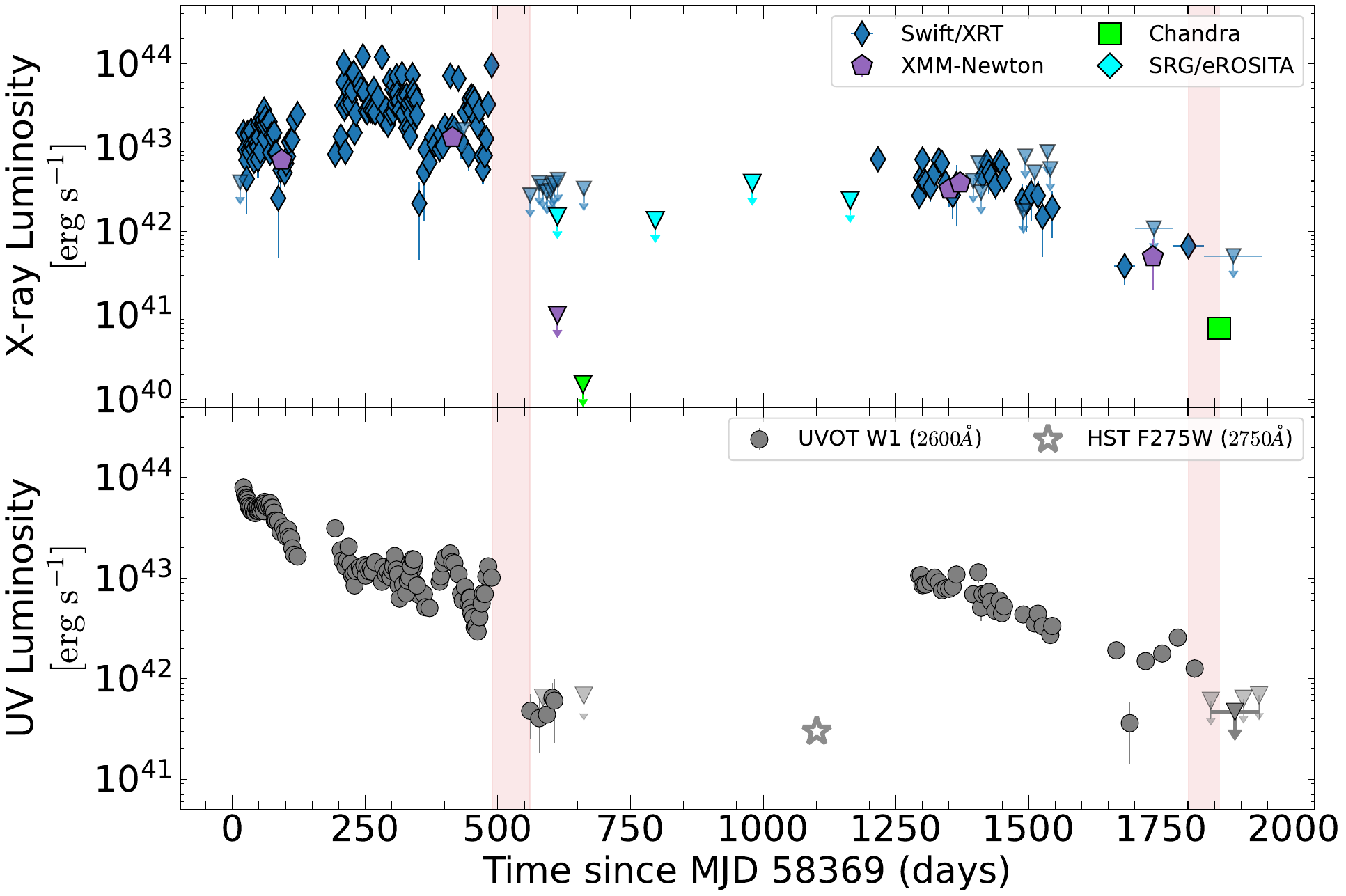}
    \caption{ {\bf Top:} \target's observed 0.3-10.0 keV X-ray luminosity evolution over the past $\sim$2000 days. The x-axis is in observer's frame. Most recent \chandra and \swift data shows a drop of $>$10 from 7$\times$10$^{42}$ to 7$\times$10$^{41}$ over two months. A similar change is also evident in the UV light curve (bottom panel). Inverted triangles represent 3$\sigma$ upper limits. This sudden change can be interpreted as a shutoff which allows us to refine the orbital period of the star that is repeatedly disrupted to be 1306$\pm$47 days. The two shutoff epochs are highlighted with red/vertical bands. The entire X-ray and UV photometry is available at \url{https://doi.org/10.5281/zenodo.10913475}.}
    \label{Fig:fig2}
\end{figure*}


\subsection{Shutoff and rebrightening times}
The first X-ray and UV shutoffs happened between days 488 and 561 while the second sharp decline occurred sometime during days 1801 and 1859 (see Fig. \ref{Fig:fig2}). These values correspond to the observation dates. Per the model of \cite{wevers3}, the orbital period of the surviving star is the time between shutoffs, which based on the above values is 1306 $\pm$ 47 days\footnote{Note that \cite{wevers3} estimated the orbital period of the star by assuming that the fallback time between the first and second encounters was the same, and while this is likely a fairly good approximation, observing the successive shutoffs is more direct. See Section \ref{sec:dis} for additional discussion.}. The uncertainty is derived from adding the range in cutoff times in quadrature. Using this we can formulate a crude ephemeris to predict the n$^{th}$ shutoff to be:
\[
    t_{\rm shutoff}^{n}(MJD) = (58893.5\pm29) + (n-1)(1306\pm47)
\]
This equation implies that the next (third) shutoff should occur sometime between 2 Jan 2027 and 17 July 2027, assuming that the star survived its second encounter. Alternatively, if the star was completely destroyed during the second encounter, then there would be no third cutoff and the luminosity would continue to smoothly decline.\footnote{We note that this possibility may provide a unique opportunity to explore the differences in emission produced by fallback from partial and complete disruption events in the same system, i.e., comprising the same black hole mass and spin and stellar orbit.}

The fallback time after the second pericenter passage of the star is the time between the first shutoff and the second rebrightening (between 1164 and 1216 days). The fallback time will differ from one encounter to the next because mass is stripped from the progenitor and the star is imparted net rotation \citep{bandopadhyay24}, and hence accurately 
predicting the next rise is not as straightforward as predicting the orbital period. However, if we assume a similar fallback time, then the next rise in flux should happen around 2495$\pm$54 days, which corresponds to an MJD 60864$\pm$54 (15 May -- 31 Aug 2025).  

The latest \chandra data point (green square in Fig. \ref{Fig:fig2}) is two orders of magnitude below the peak of the second outburst and an order of magnitude below a previous XRT detection roughly  two months earlier. However, it is possible that the latest \chandra data and the corresponding UV upper limits may be due to anomalous source variability. For this reason, we refer to this as a potential shutoff. This can be confirmed with further deep X-ray and UV observations between now and the predicted next rebrightening in 2025.

\section{Discussion and conclusions}\label{sec:dis}
The rpTDE model proposes that a star is on a highly eccentric ($0.99 \lesssim e < 1$) orbit about an SMBH, with the short orbital period and high eccentricity provided by the Hills mechanism \citep{cufari22, wevers3}. Since the fallback time inferred from the observations is $\sim$ 600-700 days, the SMBH powering the emission from \target must be large and the disruption must be partial, as both of these effects increase the return time of the debris above the $\sim (30\pm 5)\times\left(M_{\bullet}/10^6M_{\odot}\right)^{1/2}$ days that is characteristic of complete disruptions, with $M_{\bullet}$ the SMBH mass \citep{coughlin22, bandopadhyay24b}. 

When the SMBH mass is large (as is inferred to be the case for \target; 10$^{7.7\pm0.4}$; see \citealt{wevers3}), 
the accretion rate should be strongly coupled to the fallback rate of debris, because the pericenter distance is highly relativistic and the accretion timescale is short relative to the fallback time. 
This model then predicts that the accretion rate should shut off when the surviving core returns to the (partial) tidal disruption radius (\citealt{wevers3}; see also \citealt{liu23}), the reason being that the Hill sphere that separates material bound to the black hole and bound to the star grows with time approximately as $\propto t^{2/3}$, where $t$ is time since pericenter \citep{coughlin19}. Therefore, when the surviving core returns to pericenter, there is a sudden drop in the mass supply to the SMBH and the luminosity plummets. The simultaneous plummeting of the optical/UV emission alongside the X-ray is also consistent with the interpretation that the optical/UV emission is tied to X-ray emission\footnote{This seems inconsistent with the interpretation that the optical/UV is sourced from a large-scale outflow that is causally disconnected from the X-ray emission (e.g., \citealt{price24}).} that originates from the innermost few gravitational radii, which in this case may be physically produced by circularization shocks, accretion, or both (the former may also give rise to the nonthermal electrons powering the corona; cf.~\citealt{sironi24}).

The time between successive cutoffs in emission should therefore closely track the orbital period of the stellar core. Because the orbital period is related to the orbital energy and the orbital energy can at most be reduced by the binding energy of the star, there is effectively no change in this recurrence time on a per-orbit basis \citep{cufari23, bandopadhyay24}.\footnote{This holds for orbits generated by the Hills disruption of a tight binary, where by tight we mean that the binding energy of the binary is comparable to the binding energy of the captured star (which was one of the members of the original binary). In this case the binding energy of the captured star's orbit is larger than that of the star itself by a factor of $(M_\bullet/M_\star)^{1/3}$ \citep[e.g.][]{cufari22}. On the other hand, for a standard TDE in which the binding energy of the star's orbit is $\sim 0$, the change in the energy of the core during core reformation \citep[e.g.][]{nixon21,nixon22} or due to a positive-energy kick \citep[e.g.][]{manukian13,gafton15} can give rise to substantial differences in the orbital period between successive partial disruptions.} On the other hand, the time between the cutoff and the next rebrightening equals the fallback time of the most-bound debris that is related to the properties of the star and its rotation rate, and the latter changes as a consequence of the tidal interaction with the SMBH (since the imparted spin is prograde with respect to the orbital angular momentum, the result is a decrease in the return time of the debris; \citealt{golightly19}). \citet{wevers3} estimated the orbital time of the star -- and thereby predicted the time of the second cutoff -- by assuming that the fallback time was unchanged between the first and second encounter: since the first cutoff occurred at $\sim$ 500 days (since first detection) and the second brightening at $\sim$ 1200 days, the fallback time (for the second encounter) was $\sim 700$ days and the first pericenter passage must have occurred $\sim 700$ days prior to the first detection \emph{if the fallback times were identical on the first and second encounter}, making the orbital period of the star $\sim 1200$ days. If we now associate the green datapoint at day $\sim 1850$ as the second \emph{observed} cutoff (note that this date also coincides with the observed cutoff in the optical/UV, which is qualitatively in agreement with the behavior observed during the first cutoff), then this suggests that the true (i.e., from the observed successive cutoffs) orbital period of the star is between 1250-1350 days. This finding suggests that the fallback time on the second encounter was shorter than that of the first by $\sim 50-150$ days, which is different at the $\sim 10-20\%$ level. This is consistent with the theoretical results of \citet{golightly19} if the imparted spin to the star was a significant fraction of breakup (which is expected, given the importance of nonlinear interactions when the tidal field of the SMBH is comparable to the self-gravitational field of the star). If the star is spun up to a closer fraction of the angular velocity at pericenter on its third encounter, we would expect a reduced fallback time in going from the second shutoff to the start of the third electromagnetic outburst, and the observation (or lack thereof) of this feature would provide another test of this model.

A noticeable difference between the first and second outburst in AT2018fyk's lightcurve is the peak luminosity, which is reduced by a factor of $\sim 5-10$ from the first to second brightening. Under the rpTDE paradigm, this difference could be due to one or more potential factors. If, for example, the partially disrupted star was highly centrally concentrated -- which would arise naturally as a consequence of stellar evolution if the star is sufficiently massive -- then it seems plausible that the stripping of the envelope on the first encounter could leave the high-density core relatively unperturbed, the result being that the tidal radius of the surviving core is smaller. Since the pericenter distance of the star is effectively unaltered owing to the conservation of angular momentum, the $\beta$ ($=r_{\rm t}/r_{\rm p}$, with $r_{\rm p}$ the pericenter distance and $r_{\rm t}$ the tidal radius) of the encounter is reduced, resulting in less mass stripped on subsequent tidal interactions and a reduction in the accretion rate (and the luminosity). 

\citet{bandopadhyay24} recently showed (and in agreement with the suggestion by \citealt{liu23}) that more massive and evolved stars (specifically a $1.3 M_{\odot}$ and a $3 M_{\odot}$ star near the end of the main sequence) that require very deep encounters to be completely destroyed -- and are thus statistically significantly more likely to be partially disrupted -- are capable of producing flares of nearly equal amplitude after many successive outbursts, despite the fact that the amount of mass stripped from the star declines slightly per pericenter passage (see their Figures 7, 8, and 11). Contrarily, a sun-like star closer to the zero-age main sequence, which is significantly less centrally concentrated than an evolved star of the same mass, was shown by the same authors to suffer increasing degrees of mass loss per encounter, even when the pericenter distance of the encounter was a factor of $\sim 3$ larger than that required to completely destroy the star on the first encounter (see their Figure 13; see also \citealt{liu24}, who came to similar conclusions regarding the fate of a solar-like star after multiple encounters). Similarly, since a star of significantly lower mass ($\lesssim 0.1\times few M_{\odot}$) cannot evolve substantially over the age of the Universe and is effectively a $5/3$-polytrope, the range in radii where such a low-mass star could be repeatedly stripped of a small amount of mass ($\beta \equiv r_{\rm t}/r_{\rm p} \simeq 0.5-0.6$, where $r_{\rm p}$ is the pericenter distance of the star and $r_{\rm t} = R_{\star}\left(M_{\bullet}/M_{\star}\right)^{1/3}$ is the tidal disruption radius with $R_{\star}$ and $M_{\star}$ the stellar radius and mass; \citealt{guillochon13, mainetti17, miles20, cufari23}) per encounter is very fine tuned, and the detection of a second cutoff here suggests that the star must have survived at least two encounters. Thus, and despite their relative rarity, a more massive star could be the most promising candidate for producing the repeated flares in 2018fyk. A more massive star would also permit wider initial binaries (while yielding the same period of the captured star), a less relativistic pericenter distance, and a longer fallback time of the tidally stripped debris compared to the more extreme values required for a solar-like star to fit the observations (see the discussion in Section 4 of \citealt{wevers3}).

Additionally, if the Hills mechanism is responsible for placing the star on its tightly bound orbit about the SMBH, then there will be a difference between the pericenter distance of the partially disrupted (and captured) star during its initial hydrodynamical interaction between the SMBH and the (ultimately ejected) companion star and subsequent encounters. If it is such that the $\sim$ conserved pericenter distance of the captured star is larger than that of the initial interaction, then less mass will be stripped on the second encounter, resulting in a relatively smaller accretion luminosity on the second outburst. Depending on how centrally concentrated the star is and the $\beta$ of the encounter, subsequent outbursts could be progressively more or less luminous with time. As also discussed in \citet{bandopadhyay24}, the tight binary required to populate the star on its $\sim 1300$-day orbit implies that the rotation rate of the captured star is a significant fraction of the angular velocity at pericenter, which will also have a significant impact on the magnitude of successive flares and the return time of the debris \citep{golightly19}).

From the rpTDE model, the orbital time of the star is $\sim 1200-1400$ days and the fallback time is $\sim 600-800$ days (note that such long fallback times require the event to be a partial disruption; \citealt{bandopadhyay24}), which predicts that the freshly stripped debris generated on the third encounter (i.e., the second dimming around the green point in Figure \ref{Fig:fig2}) should produce a third brightening at day $\sim 2500$ post-initial-detection {(though, as noted above, the additional imparted spin to the star near pericenter could yield a time closer to day $\sim 2400$)}. This third brightening should then occur in early-2025, which is consistent with the predictions in \citet{wevers3}, and the future detection or non-detection thereof would provide strong evidence in support of or against this model. 

An alternative interpretation, as proposed by \citet{wen24}, is that the reduction in the luminosity of AT2018fyk is due to the presence of a companion black hole, and that the disrupting black hole was of very low mass compared to the primary. In such extreme-mass-ratio systems, there could be a dramatic dimming when the tidally stripped debris nears the Hill sphere of the secondary (disrupting) black hole after the first encounter, provided that the orientation of the binary is favorable \citep{coughlin18}. \citet{wen24} then proposed that the second outburst arose from accretion onto the primary. However, when the secondary is the disrupting SMBH and the stream is relatively confined to the orbital plane of the binary -- which must be the case if accretion onto the primary is responsible for the second outburst -- the distribution of the debris is highly stochastic (see, e.g., Figures 1 \& 2 of \citealt{coughlin18}, or Figures 9 \& 10 of \citealt{coughlin17}), and it is difficult to see why this scenario would produce repeated and dramatic dimmings on this same timescale.

Finally, the partial TDE results in the production of two tails of stellar debris. In typical TDEs where the star is on a parabolic orbit, the second tail is ejected from the system and yields no observational signature (with the possible exception of radio emission in the presence of circumnuclear gas; e.g., \citealt{guillochon16, yalinewich19}). However, and as noted in \citet{wevers3}, the bound nature of the stellar orbit in this case implies that the second tail may be ``less bound'' rather than unbound, with the specific energy of the second tail a function of the specific energy of the core and the energy spread imparted by the tidal field. It could be that deeper X-ray monitoring of AT2018fyk during the second (current) shutoff phase would reveal the presence of low-level emission from this second tail, and since the energy of the core is constrained from the observed orbital period, this emission would yield additional information about the properties of the star and the SMBH. rpTDEs are thus unique in their ability to more directly constrain stellar and SMBH properties in distant galaxies.  
\newline
\newline
D.R.P was supported by NASA \xmm guest observed program, proposal number 92395 (award number 035212-00001), and \chandra program 25700383 (award number 035385-00001). E.R.C.~and A.B.~acknowledge support from NASA through the \emph{Neil Gehrels Swift} Guest Investigator Program, proposal number 1922148. E.R.C.~acknowledges additional support from the National Science Foundation through grant AST-2006684, and from NASA through the Astrophysics Theory Program, grant 80NSSC24K0897. C.J.N.~acknowledges support from the Science and Technology Facilities Council (grant No. ST/Y000544/1) and from the Leverhulme Trust (grant No. RPG-2021-380). This research was supported in part by grant NSF PHY-2309135 to the Kavli Institute for Theoretical Physics (KITP).

\clearpage

\newpage

\begin{longrotatetable}
\begin{deluxetable*}{lllllllllllll}
\tablecaption{Summary of \xmm and \chandra X-ray energy spectral modeling. $^{\dagger}$The net exposure after filtering for background flares. Total exposure before correcting for flares is shown in brackets. $^{\dagger\dagger}$Net count rate (background-corrected) in the bandpass where the source is above the background. $^{\dagger\dagger\dagger}$Bandpass where the source and the background spectrum crossover. This is different for each spectrum and modeling was performed in this custom band depending on the observation. We repeated the entire analysis in a fixed 0.3-1.5 keV band and the resulting values were consistent with those reported in this table. $^{*}$Normalization of the best-fit disk black body. $^{**}$Powerlaw index. $^{\S}$Normalization value of the powerlaw model component. $^{\S\S}$Observed flux and luminosities in the 0.3-10.0 keV band. {\it tbabs*zashift(diskbb+pow)} was used for modeling. In cases where kT is indicated by ``..." a disk component was not necessary. $^{\S\S\S}$Both the {\it Chandra} spectra were fit together, hence the same spectral parameters. The powerlaw index was fixed at the best-fit \xmm value from ObsID 0921510101. \label{table}}
\tablewidth{500pt}
\tabletypesize{\scriptsize}
\tablehead{
\colhead{Telescope} & \colhead{ObsID} & \colhead{MJD} & \colhead{Exposure$^{\dagger}$} & \colhead{Count rate$^{\dagger\dagger}$} & \colhead{Bandpass$^{\dagger\dagger\dagger}$} & \colhead{kT} & \colhead{N$_{kT}$$^{*}$} & \colhead{$\Gamma$$^{**}$} & \colhead{N$_{\Gamma}^{\S}$} & \colhead{Flux$^{\S\S}$$\times$10$^{-13}$} & \colhead{Luminosity$^{\S\S}$} & \colhead{C-stat/dof} \\ 
\colhead{} & \colhead{} & \colhead{} & \colhead{(ks)} & \colhead{ (counts/sec)} & \colhead{(keV)} & \colhead{(keV)} & \colhead{} & \colhead{} & \colhead{($\times$10$^{-5}$)} & \colhead{(\cgs)} & \colhead{(10$^{42}$ erg s$^{-1}$)} & \colhead{}
} 
\startdata\\
XMM & 0831790201 & 58461.72 & 17.0 (33) & 0.403$\pm$0.004 & 0.3-2.5 & 0.123$^{+0.005}_{-0.004}$ & 493$^{+100}_{-164}$ & 3.41$^{+0.68}_{-1.38}$ & 3.4$^{+1.4}_{-1.9}$ & 7.7$^{+0.2}_{-0.1}$ & 7.0$^{+0.3}_{-0.1}$ & 42.4/48 \\
\\
XMM & 0853980201 & 58783.33 & 35.0 (55) & 0.687$\pm$0.004 & 0.3-9.0 & 0.146$^{+0.006}_{-0.006}$ & 170$^{+32}_{-25}$ & 2.07$^{+0.06}_{-0.06}$ & 22.6$^{+1.4}_{-1.4}$ & 15.7$^{+0.1}_{-0.2}$ & 13.3$^{+0.1}_{-0.2}$ & 192.0/155 \\
\\
XMM & 0911790601 & 59719.89 & 8.2 (29) & 0.175$\pm$0.005 & 0.3-5.0 & 0.095$^{+0.035}_{-0.032}$ & 211$^{+1700}_{-173}$ & 2.35$^{+0.19}_{-0.22}$ & 8.6$^{+0.8}_{-1.1}$ & 3.7$^{+0.3}_{-0.3}$ & 3.2$^{+0.3}_{-0.3}$ & 72.1/82 \\
\\
XMM & 0911791401 & 59739.84 &  3.1 (10.6) & 0.130$\pm$0.009 & 0.3-0.8 & \nodata & \nodata & 2.77$^{+0.54}_{-0.54}$ & 11.1$^{+4.4}_{-3.4}$ & 4.4$^{+1.4}_{-0.6}$ & 3.7$^{+0.8}_{-0.6}$ & 15.0/11 \\
\\
XMM & 0921510101 & 60102.76 &  8.8 (43.1) & 0.011$\pm$0.001 & 0.3-1.0 & \nodata & \nodata & 1.96$^{+0.86}_{-0.88}$ & 1.1$^{+0.6}_{-0.4}$ & 0.6$^{+0.7}_{-0.2}$ & 0.5$^{+0.3}_{-0.2}$ & 19.1/16 \\
\\
Chandra$^{\S\S\S}$ & 28294 & 60227.71 &  32.6 (33) & (3.8$\pm$1.2)$\times$10$^{-4}$ & 0.5-7.0 & \nodata & \nodata & 1.96 & 0.18$^{+0.08}_{-0.06}$ & 0.09$^{+0.01}_{-0.02}$ & 0.07$^{+0.02}_{-0.01}$ & 38.5/56 \\
\\
Chandra$^{\S\S\S}$ & 28972 & 60228.58 &  19.8 (20) & (5.0$\pm$1.7)$\times$10$^{-4}$ & 0.5-7.0 & \nodata & \nodata & 1.96 & 0.18$^{+0.08}_{-0.06}$ & 0.09$^{+0.01}_{-0.02}$ & 0.07$^{+0.02}_{-0.01}$ & '' \\
\\
\enddata
\end{deluxetable*}
\end{longrotatetable}

\newpage
\bibliographystyle{aasjournal}
\bibliography{refs}

\end{document}